\documentclass[fdp,fleqn]{w-art}
\usepackage{times,amssymb}
\usepackage{w-thm}

\usepackage[]{graphicx}
\begin{document}

\DOIsuffix{theDOIsuffix}
\Volume{xx}
\Issue{x}
\Month{xx}
\Year{xxxx}

\pagespan{1}{}
\Receiveddate{xxx}
\Reviseddate{xxx}
\Accepteddate{xxx}
\Dateposted{xxx}


\title[Lifshitz Black Holes]{Asymptotically Lifshitz Black Holes in Einstein-Maxwell-Dilaton Theories}

\author[J. Tarrio]{Javier Tarrio}

\address[]{Institute for Theoretical Physics and Spinoza Institute, Universiteit Utrecht, 3584 CE, Utrecht, The Netherlands. \\ E-mail:~\textsf{l.j.tarriobarreiro@uu.nl}}

\begin{abstract}

We study  Einstein-Maxwell-dilaton theories with a cosmological constant and $U(1)^N$ gauge symmetry, considering metrics asymptotically approaching the Lifshiftz metric. 
We  study the dependence of the phase diagram on the value of the dynamical exponent. Along the way, we apply holographic renormalization and propose a counterterm valid for arbitrary dimension and dynamical exponent in our setup.

\end{abstract}

\maketitle                   

\section{Context and setup}

This note is concerned with the study of black hole solutions to the Einstein-Maxwell-Dilaton system with negative cosmological constant, and in particular solutions in which the time and the space coordinates scale differently:
$
t \to \lambda^z t$,  $x^i \to \lambda \, x^i \ ,
$
where $z$ is called the dynamical exponent. This kind of scaling is observed in many condensed matter systems, and has been investigated with great detail from a holographic perspective since the seminal work \cite{Kachru:2008yh}, in which one includes an extra holographic coordinate, scaling as $r \to \lambda^{-1} r$, with a boundary in which the field theory lives. The precise relation between the field theory of interest and the dual holographic theory is not yet known, and a construction of the holographic dictionary for this theories is under current investigation   \cite{Ross:2011gu,Baggio:2011cp,Mann:2011hg}.

In \cite{Kachru:2008yh}, the authors proposed that gravitational theories  realizing the Lifshitz symmetry should asymptotically approach the so-called Lifshitz metric
\begin{equation} \label{eq.lifmet}
\mathrm{d}s^2 =  \ell^2\frac{\mathrm{d} r^2}{r^2} - \frac{r^{2 z}}{\ell^{2z}} \mathrm{d} t^2 + \frac{r^2}{\ell^2} \mathrm{d} \vec x_{d-1}^2  \ ,
\end{equation}
which has $1+d(d+1)/2$ Killing vectors, related to the generators of the symmetry algebra in the field theory (spatial rotations, spatial translations, time translations and dilatations) \cite{Hartnoll:2009sz}.
Such a metric can be obtained in Einstein gravity only in the presence of matter fields. Most of the solutions constructed in the literature consider a system in which gravity is coupled to a massive vector field, described by the Proca action. If the mass of the vector field has the appropriate value the  metric is asymptotically Lifshitz. The caveat with the Proca approach to asymptotically Lifshitz black holes is that only a few solutions are known analytically (and for specific values of dimensions, $d$, and the dynamical exponent $z$), and in top-down approaches the solution relies on numeric calculations \cite{Amado:2011nd}. To find a generic solution for $d>2$ and $z\geq 1$ it has proven useful to consider instead the Einstein-Maxwell-Dilaton action with negative cosmological constant \cite{Taylor:2008tg}.

Following the holographic philosophy, the construction of black hole setups asymptotically approaching  \eqref{eq.lifmet} are interesting to describe thermal field theories. Furthermore, if the effects of a global $U(1)$ symmetry in the field theory are to be included, one has to consider a gauge field in the bulk of the $d+1$-dimensional gravitational theory. With this in mind we study the action \cite{Tarrio:2011de}
\begin{equation}\label{eq.action}
S= - \frac{1}{16\pi G_{d+1}} \int_M \mathrm{d}^{d+1}x\, \sqrt{-g} \left[  R -2 \Lambda - \frac{1}{2} \left( \partial \phi \right)^2 - \frac{1}{4} \sum_{i=1}^N e^{\lambda_i \phi} F_i^2 \right] +S_{GH} \ ,
\end{equation}
where the second term is the Gibbons-Hawking term, needed to have a well defined variational problem for the graviton. Below we will see that additional boundary terms must be added to obtain a finite on-shell action. These terms and the GH term do not affect the equations of motion derived from the action, which present a solution describing a charged black hole/brane
\begin{eqnarray}
\mathrm{d} s^2 & = & \frac{\ell^2}{b_k}\frac{\mathrm{d} r^2}{r^2} - \frac{r^{2 z}}{\ell^{2z}}b_k\, \mathrm{d} t^2 + r^2 \mathrm{d} \Omega_{k,d-1}^2 \ , \label{eq.solmetric}\\
b_k & = &1+ k \left( \frac{d-2}{d+z -3} \right)^2 \frac{\ell^2}{r^2}- \frac{m}{r^{d+z-1}}+\sum_{j=2}^{N-1}\frac{\rho_j^2\, \mu^{-\sqrt{2\frac{z-1}{d-1}}}\ell^{2 z}}{2(d-1)(d+z-3)}  \frac{1}{r^{2(d+z-2)}} \ , \label{eq.solbk}\\
A_{1,t}' & = & \ell^{-z}  \sqrt{2 (d+z-1)(z -1)} \, \mu^{\sqrt{\frac{d-1}{2(z-1)}}}\, r ^{d +z -2} \ , \label{eq.solAt1}\\
A_{j,t}' & = & \rho_j\, \mu^{-\sqrt{2\frac{z-1}{d-1}}} \,r ^{2-d -z} \ , \qquad \qquad \qquad ( j=2,\cdots,N-1) \,\label{eq.solAt2}\\
A _{N,t}' & = & \ell^{1-z}  \frac{\sqrt{2k(d -1) (d -2) (z -1)} }{\sqrt{d +z -3}}  \mu^\frac{(d-2)}{\sqrt{2(d-1)(z-1)}}\, r ^{d +z -4}\ , \label{eq.solAt3}\\
e^\phi & = & \mu \,r^{\sqrt{2(d-1)(z-1)}} \ , \label{eq.solscalar}
\end{eqnarray}
where we have set $\Lambda = -(d+z-1)(d+z-2)/2\ell^2$ and $\mu$ is the amplitude of the scalar field. Notice that in the metric we have generalized the spatial section to admit $d-1$-dimensional maximally symmetric manifolds with positive, null or negative curvatures ($k=1,0,-1$ respectively). In the case $k=-1$   one of the gauge fields acquires a phase and we do not consider it further. The case $k=0$ can be obtained by taking the large-sphere-radius limit from the $k=1$ case \cite{Chamblin:1999tk}. We will focus in the last case in the following, keeping $k$ explicit in some expressions. From now on we will set $N=3$ for concreteness. In the $z\to1$ limit, in which \eqref{eq.lifmet} goes to the $AdS_{d+1}$ metric, this solution goes to the AdS-Reissner-Nordstr\"om solution. This model has been used to study the conditions that strongly interacting fermions should obey to satisfy the ARPES sum rules  \cite{Gursoy:2011gz}.

The caveat this solution presents is the behavior of the $A_{{1,3}}$ gauge fields and the scalar $\phi$. These diverge at the boundary, and the variation  relative to the Killing vector $\xi_D=z t\partial_t - r \partial_r + x^{i} \partial_{i}$ is not vanishing, therefore spoiling the asymptotic Lifshitz symmetry algebra. We will not try to make sense of these fields in this note, and therefore the model we use must not be considered as a dual to a non-relativistic field theory, but as an IR effective theory that must be completed in the UV.

From the solutions \eqref{eq.solAt1} and \eqref{eq.solAt3} we can define charges associated to the badly behaved radial electric fields (see equation \eqref{eq.charge} below). These charges are fixed in terms of $d$, $z$ and $\mu$ (contrary to the case in equation \eqref{eq.solAt2}, where it is given by a free parameter $\rho_2\equiv \rho$). This  suggests that these two fields should be considered in an ensemble in which their charges are kept fixed. The original Einstein-Maxwell-Dilaton action is appropriate for an ensemble in which the charges can vary \cite{Chamblin:1999tk}, and therefore it must be modified. In other words, the boundary condition for the gauge fields imposed at the boundary of  spacetime is not a Dirichlet boundary condition, but Neumann. To account for this we have to supplement the action with a boundary term to make the variational problem well defined. This is accounted by
\begin{equation}\label{legendre}
\tilde S= S - \frac{1}{16\pi G_{d+1}}\int_{\partial M} \mathrm{d}^{d}x \sqrt{-h} \, n_\mu \left(\frac{1}{2} e^{\lambda_1 \phi} A_{1,\nu} F^{\mu\nu}_{1} +\frac{1}{2} e^{\lambda_3 \phi} A_{3,\nu} F^{\mu\nu}_{3} \right)\ ,
\end{equation}
where $h_{mn}$ is the boundary metric and $n$ is a unit vector orthogonal to $r=\text{constant}$ hypersurfaces.

An alternative (equivalent) course of action is to perform a Legendre transform of the action with respect to the electric fields under consideration \cite{McGreevy:2009xe}. This transformation changes the equations of motion, but \eqref{eq.solmetric}-\eqref{eq.solscalar} is still  a solution, only that the fields $A_{t,1}$ and $A_{t,3}$ are not  present in the new setup, since the solution has been already plugged into the Legendre transformed action.

\section{Holographic renormalization}

\paragraph{On-shell action}

As usual in holography, the evaluation of the on-shell action \eqref{legendre} diverges when one evaluates it at the boundary, and a renormalization procedure is needed. Since the publication of the solution \eqref{eq.solmetric}-\eqref{eq.solscalar}, several papers appeared dealing with the holographic renormalization of asymptotically Lifshitz spacetimes \cite{Ross:2011gu,Baggio:2011cp,Mann:2011hg}. These, however, focus on the Proca action, and  we will not consider them in the following. Indeed, the renormalization we describe in this section can be understood from the procedure for asymptotically AdS spacetimes with a few changes.

To construct the needed counterterm action in asymptotically $AdS$ spacetimes one has to solve a Hamilton-Jacobi equation \cite{Balasubramanian:1999re}, which can be done systematically by expanding the new term in a derivatives series. For a counterterm depending only on gravitational terms this accounts to an expansion in powers of the boundary Ricci scalar $R_{(h)}$, contractions of the boundary Ricci tensor $R_{(h),mn}R_{(h)}^{mn}$, the boundary Riemann tensor, and generic combinations of them. Here we denote the boundary metric as $h_{mn}$. The number of counterterms to add depends on the dimensionality of the theory under consideration. Generically, terms with $R_{(h)}^\alpha$ (and combinations of $\alpha$ boundary Ricci scalars, Ricci tensors and Riemann tensors with all indices contracted) will be needed if the field theory dimension is $d=2+2\alpha$.

 In the present case such terms will regularize a field theory dimensionality $d=3+2\alpha-z$. This implies that to regularize the theory under consideration for generic $z$ we need to consider an infinite number of counterterms. This is an impossible task, since for increasing value of $\alpha$ the combinations of curvature quantities increase in number (see \cite{Kraus:1999di} for the generic expression up to sixth order in derivatives in asymptotically AdS spacetimes).
 
There is one property of our solution we can use to circumvent this, though. Focusing in the $k=1$ case, the boundary metric has a $\mathbb{R}_t\times S^{d-1}$ topology and is static, implying that the boundary Riemann tensor takes a very simple form: any component with an index along the $t$ direction vanishes, and the spatial part is that of a maximally symmetric $d-1$-dimensional manifold with positive curvature. In particular, this means that any contraction of $\alpha$ boundary Riemann tensors, Ricci tensors and Ricci scalars will be proportional to $R_{(h)}^\alpha$, implying that the counterterm, expressed partially on-shell, is a power series in $R_{(h)}$.

For fixed $d$ and $z$ we can truncate the $R_{(h)}$ series and find what are the coefficients weighting each power of the boundary Ricci scalar to cancel divergences. When several examples are worked out, we can  find a generic expression for the coefficient in front of every power of the  boundary Ricci scalar,  for arbitrary $d$ and $z$. Assuming that these coefficients are the first coefficients of the infinite series in $R_{(h)}$, we induce the form of the $n^{th}$ coefficient accompanying $R_{(h)}^n$ in the series, and re-sum it. After this is done we find the counterterm\footnote{This counterterm in the asymptotically AdS case, $z=1$, was suggested in \cite{Mann:1999pc} for $d=3$, and in  \cite{Kraus:1999di} it was written for generic $d$.}
\begin{equation}\label{eq.ctterm}
S_{ct} = - \frac{1}{8\pi G_{d+1}}\int_{\partial M} \mathrm{d}^d x  \sqrt{-h} \frac{d-1}{\ell}  \sqrt{1+ \frac{(d-2)\ell^2 R_{(h)}}{(d-1)(d+z-3)^2}} \ .
\end{equation}
This counterterm precisely cancels the value of the on-shell action \eqref{legendre} in the case where no black hole is present, meaning that we could have defined it by covariantizing the on-shell action (using he boundary Ricci scalar) in the zero temperature, zero charge setup. This is how the $z=1$ case was found in  \cite{Kraus:1999di}, and shows that the counterterm is valid  in any number of dimensions and for arbitrary dynamical exponent $z\geq 1$, despite it being derived in the inductive way described before.

For the thermal case we obtain a finite renormalized on-shell action, $\tilde S_{ren}=\tilde S_{on-shell}+S_{GH}+S_{ct}$, evaluated at a cutoff (and taken to the boundary)
\begin{equation}\label{grandcan}
\tilde S_{ren} = \frac{\beta V_{d-1}}{16\pi G_{d+1}\ell^{1+z}} \left(m(z-2) -2 (z-1) r_h^{d+z-1}  -  \frac{2(d-2)^2 (z-2) \ell^2 r_h^{d+z-3} }{(d+z-3)^2} \right) \ .
\end{equation}

One could have also considered the Legendre transformed action with respect to the physical charge $\rho$. It can be checked easily that the same counterterm does the job in this case, since the Legendre transform is equivalent to the inclusion of a boundary term (see equation \eqref{legendre} and the comments around it), which in our case give a finite contribution. After this consideration, the on-shell Legendre transformed action reads 
\begin{equation}\label{can}
\tilde {\tilde S}_{ren} = \frac{\beta V_{d-1}}{16\pi G_{d+1}\ell^{1+z}} \left(m(2d+z-4) -2 (d+z-2) r_h^{d+z-1}  -  \frac{2(d-2)^2 \ell^2 r_h^{d+z-3} }{d+z-3} \right) \ .
\end{equation}

\paragraph{Internal energy}

The trick used in the renormalization of the on-shell action is not useful when we try to renormalize the stress-energy tensor in the field theory. The reason is that we have used the fact that topology of the boundary metric is $\mathbb{R}_t\times S^{d-1}$ to express curvature invariants composed by the Riemann tensor, the Ricci tensor and the Ricci scalar as powers of the Ricci scalar alone. However, the calculation of the boundary stress energy tensor involves a variation of the action with respect to the boundary metric, and this result will be sensitive to the precise combination of curvature tensors that we have. Despite these considerations, let us write here the Brown-York tensor as derived by considering the counterterm \eqref{eq.ctterm}
\begin{equation}\label{eq.setensor}
T_{mn} = \frac{1}{8\pi G_{d+1}} \left( K_{mn} - K \, h_{mn} +  \frac{d-1}{\ell} \sqrt{ 1 + \frac{(d-2)\ell^2 R_{(h)} }{(d-1)(d+z-3)^2} } h_{mn}  \right) \ .
\end{equation}
In principle, $T_{mn}$ should have corrections involving terms proportional to contractions of curvature tensors with two free indices (for example $R_{{(h),mn}}$ or $R_{(h),pmqn}R_{(h)}^{pq}$).  As an example, the lengthy expression  one gets considering the most generic counterterm up to six derivatives of the boundary metric can be seen in \cite{Das:2000cu}.

Now, the staticity of our solution implies that any component of the Riemann tensor with an index in the time direction vanishes. This in turn means that the $tt$ component can still be evaluated, since the corrections to \eqref{eq.setensor}  cancel for this component. With the evaluation of $T_{{tt}}$ we can calculate the renormalized energy density in the field theory straightforwardly, obtaining
\begin{equation}
E = \int d^{d-1}x \sqrt{\det h_{ij}} k^m \xi^n T_{mn} = \frac{V_{d-1}}{16\pi G_{d+1}} \frac{m (d-1)}{\ell^{1+z}} \ ,
\end{equation}
with $\det h_{ij}$ involving only the spatial direction components of the boundary metric, $\xi=\partial_t$ a Killing vector and $k$ the unit vector orthogonal to $t=\text{constant}$ surfaces. This result coincides withe the Komar mass result used in \cite{Tarrio:2011de} by subtracting a reference background.

\section{Thermodynamics}

The temperature can be defined from the inverse of the euclidean time by the absence of singularities when performing a Wick rotation, giving an explicit expression $T=T(r_h, \rho, \mu)$ that can be found in \cite{Tarrio:2011de}. The mass parameter $m\geq0$ can be traded by the radius of the horizon, which is defined by $b_k(r_h)=0$.  The entropy is given, as usual, by the Bekenstein-Hawking formula
$
S = V_{d-1}  r_h^{d-1}/4 G_{d+1} 
$.

The charge carried by the gauge field $A_2$ can be obtained from the Gauss law, and the chemical potential associated to this charge in the field theory is obtained, as usual in the holographic context, from the value of the corresponding gauge field at the boundary. These calculations give
\begin{equation}
   Q  =  \frac{1}{16\pi G_{d+1}} \int e^{\lambda_2 \phi}*F_2 = \frac{V_{d-1} \ell^{z-1}\rho}{16\pi G_{d+1}}\ ,  \qquad  \Phi=A_{2,t}(\infty)  =   \frac{\rho\, \mu^{-\sqrt{2\frac{z-1}{d-1}}} }{d+z-3} r_h^{3-d-z}  \ ,\label{eq.charge}
\end{equation}
where we have used the usual boundary condition $A_{2,t}(r_h)=0$.

An analogous calculation for the chemical potential associated to the badly behaved fields $A_1$ and $A_3$ implies a divergent value. However, as we work in an ensemble in which the charge associated to these fields is kept fixed, explicit factors of these chemical potentials will not appear in the thermodynamic relations.

\paragraph{Thermodynamic potentials}

With the relations given above it is straightforward to check the relation
\begin{equation}
W= T I_{ren}  = E - TS - Q \Phi \ ,
\end{equation}
with $T$ the temperature. Actually, the differential thermodynamic relations in the grand-canonical ensemble (in which $Q$ is allowed to vary and the potential $\Phi$ is fixed) are also satisfied. We then conclude that the renormalized on-shell action \eqref{grandcan} (times the temperature) corresponds to the free energy of the field theory in the grand-canonical ensemble. 

For the Legendre transformed action \eqref{can} one finds
\begin{equation}\label{canonicalthermo}
F=T (\tilde I_{ren}-\tilde I_{ren}^{T=0} )= \Delta E - TS  \ ,
\end{equation}
and looking at differential relations the identification with the free energy in the canonical ensemble is realized. This confirms the educated guess  in \cite{Tarrio:2011de}, where these potentials were defined from the r.h.s. of the previous two equations, without going through the renormalization procedure. In \eqref{canonicalthermo} we have subtracted the $T=0$ value to make contact with the analysis in that paper. Here $\Delta E=E-E_{ext}$ is the energy with the value corresponding to the extremal solution subtracted.

\paragraph{Phase diagrams}

\begin{figure}[htb]
\begin{center}
\begin{tabular}{ccc}
 \includegraphics[scale=0.35]{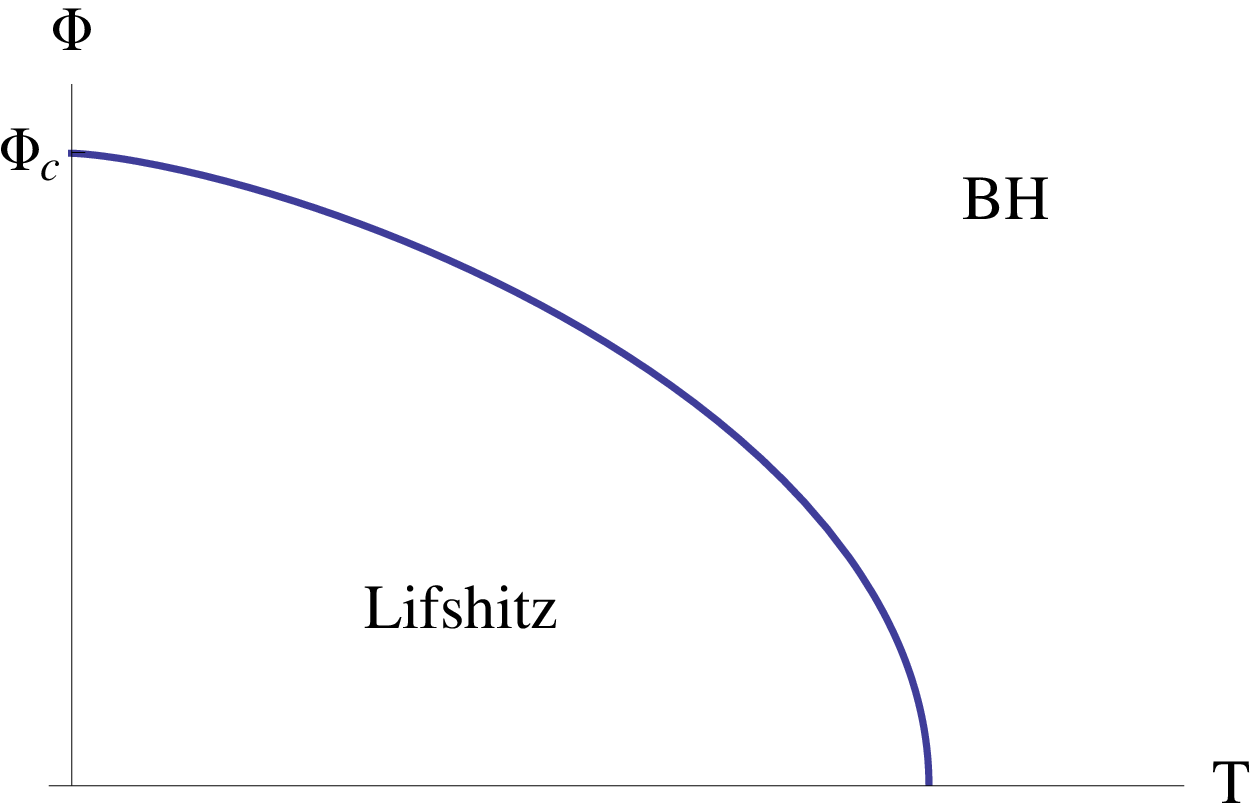} & \includegraphics[scale=0.35]{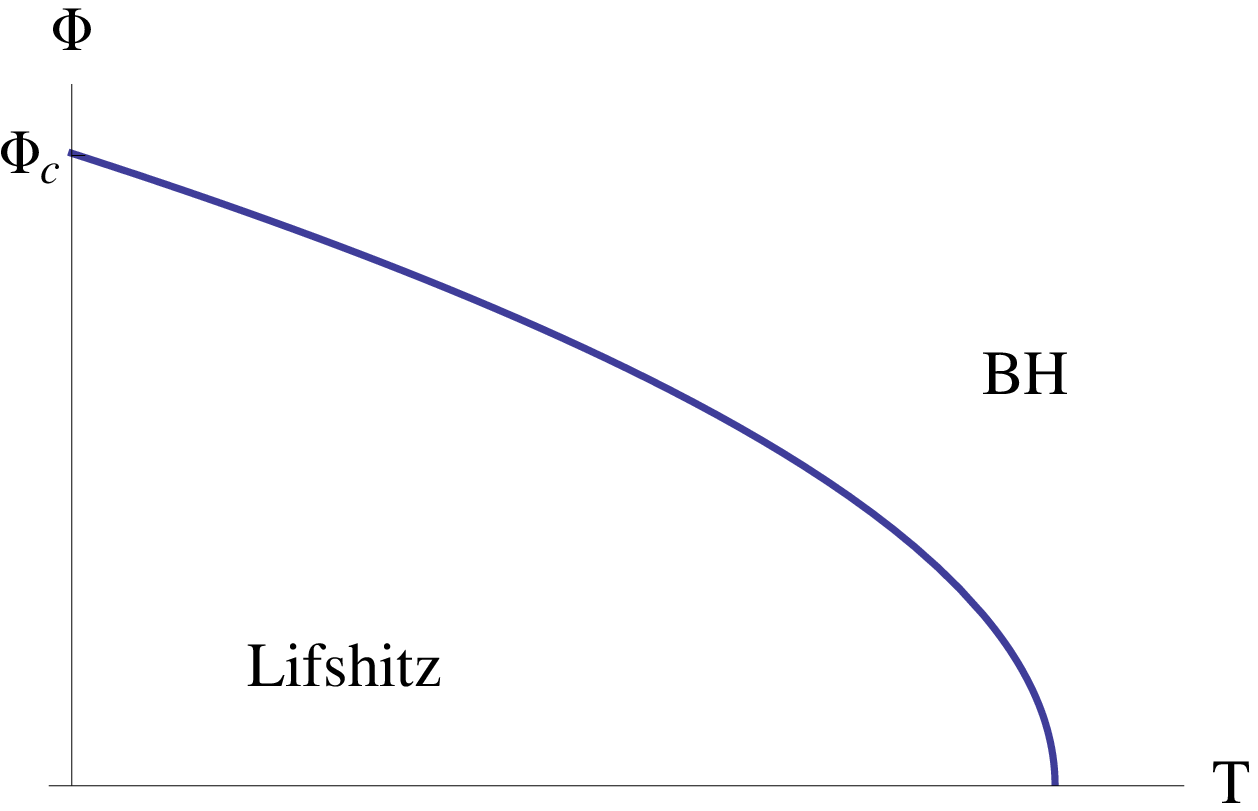} & \includegraphics[scale=0.35]{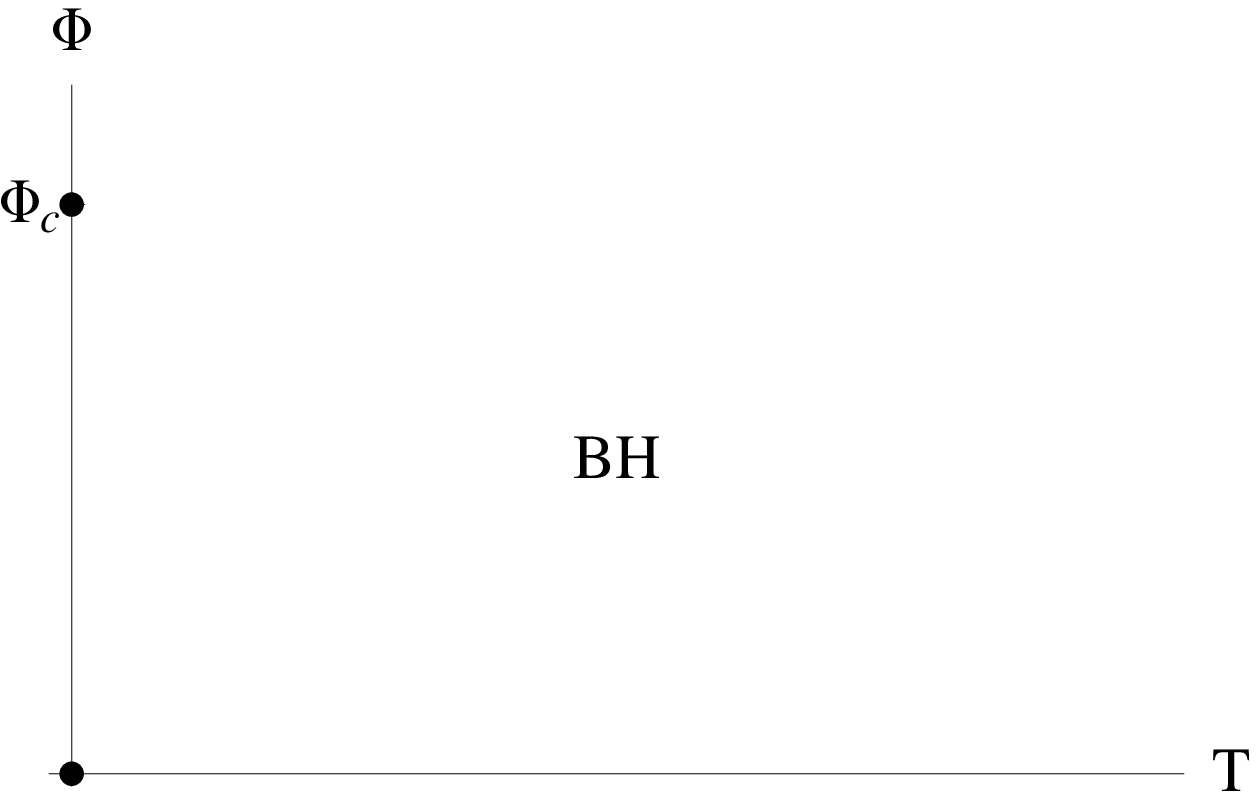}  \\ 
\includegraphics[scale=0.35]{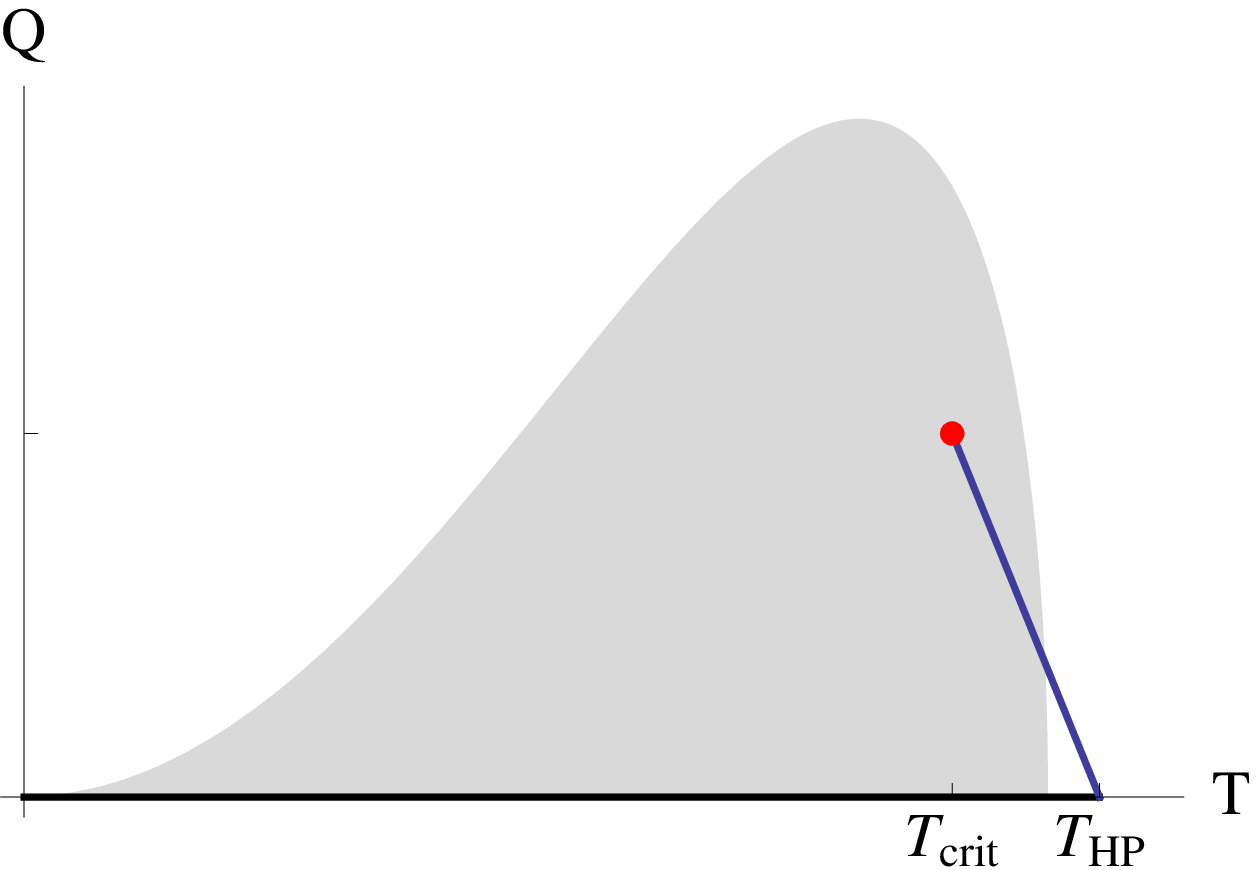} &  \includegraphics[scale=0.35]{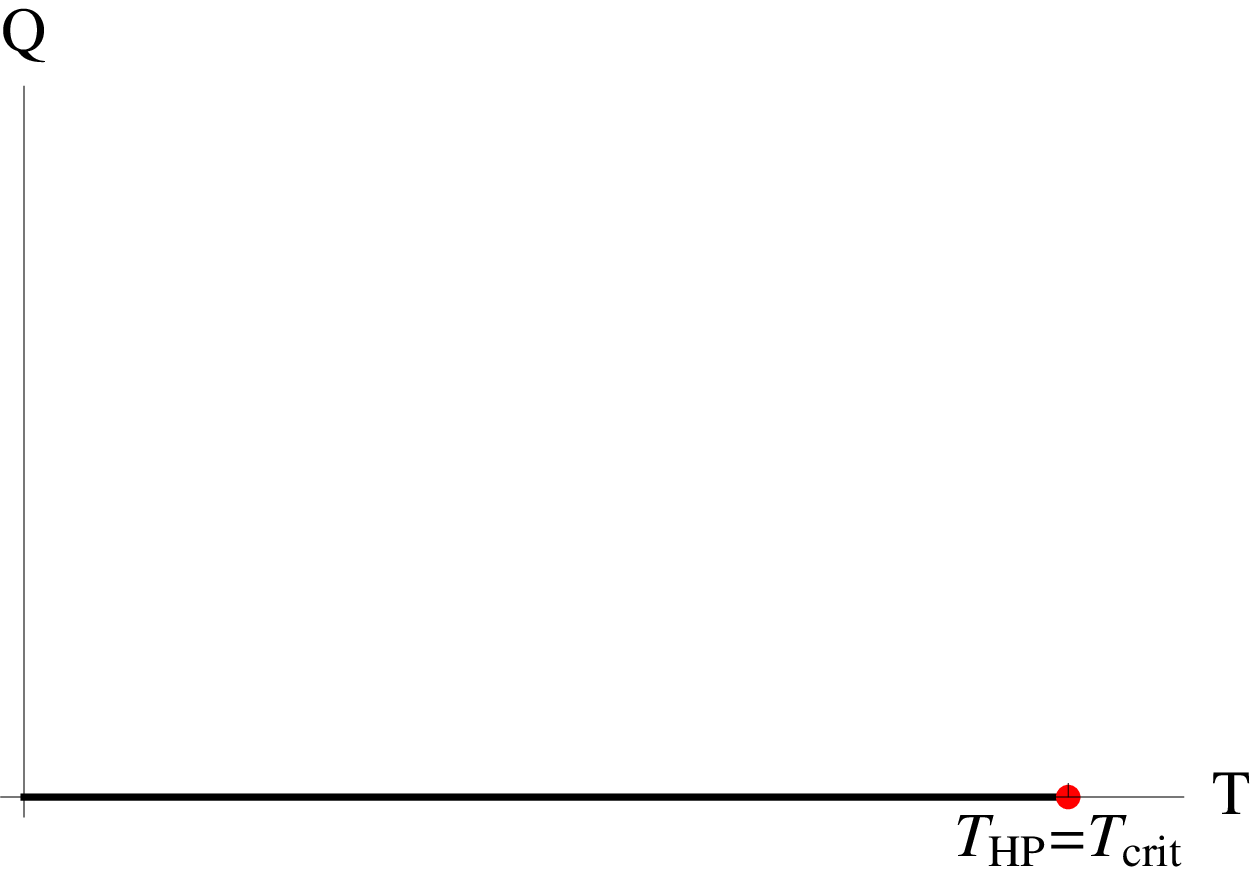}  & \includegraphics[scale=0.35]{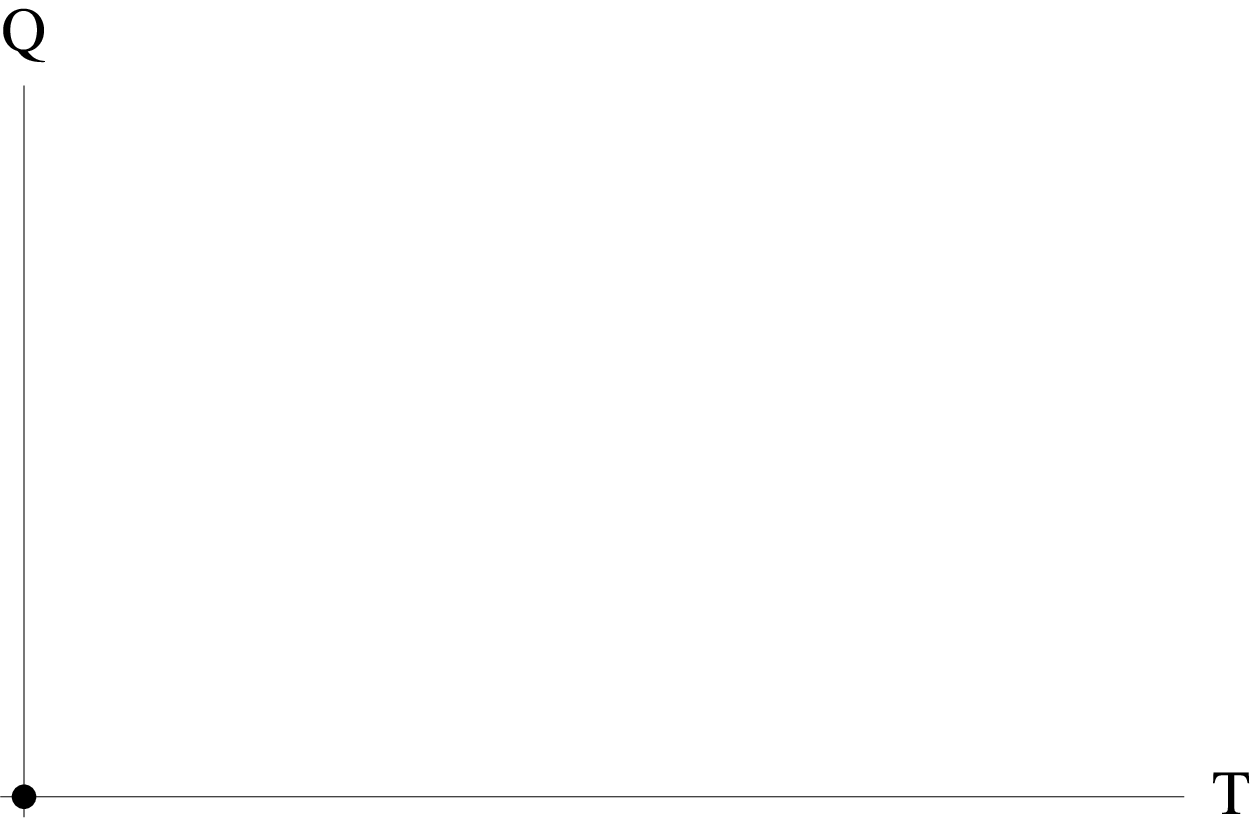} \\
\end{tabular}
\caption{Sketch of the phase diagrams obtained here for different values of the dynamic exponent $z<2$ (left), $z>2$ (right) and $z=2$ (middle) in the grand-canonical (top) and canonical (bottom) ensembles.}\label{fig.phases}
\end{center}
\end{figure}

We show the dependence of the phase diagrams with the dynamical exponent in figure \ref{fig.phases}, for both the canonical and grand-canonical ensembles. 

For $z<2$ the situation is analogous to the AdS-Reissner-Nordstr\"om case discussed in \cite{Chamblin:1999tk} (which corresponds to the $z=1$ case in this note). There is a line of first order phase transitions in the grand-canonical ensemble separating the thermal Lifshitz solution (no black hole but periodic euclidean time) valid at low values of the temperature and the chemical potential, from the black hole setup. For large values of the chemical potential and vanishing temperature the dominating solution is a finite-entropy extremal black hole, which is expected to decay into an unspecified ground state. In the canonical ensemble there is a line of first order phase transitions between small and large black holes, terminating at a critical point above which we have a crossover. At $Q=0$ we have the analogous to the Hawking-Page transition \cite{Hawking:1982dh}. The shadowed region in this diagram marks electrically unstable setups.

For $z=2$ the situation in the grand-canonical case stays unchanged, but in the canonical case the line of first order phase transitions gets reduced to just the critical point at $Q=0$, \emph{i.e.}, to the Hawking-Page transition. The electric instability also disappears for this value of the dynamical exponent.

Finally, for $z>2$ the phase diagrams simplifies in both ensembles. In the grand-canonical one the first order transitions disappear and the black hole setup (extremal black hole setup in the $T=0$ case) dominates everywhere in the phase diagram, except at a critical value of the chemical potential (for $T=0$) where the dominating solution is Lifshitz spacetime. In the canonical ensemble we do not find any phase transition.

\section{Conclusions}

We have studied the thermodynamics of the solution presented in \cite{Tarrio:2011de} using holographic renormalization to calculate the free and internal energies of the system, related to the on-shell action and the $tt$ component of the boundary Brown York tensor, respectively. 

The fact that a counterterm can be given for generic values of $d$ and $z$ is possible due to the extremely simple boundary topology of the solution, which allows to re-sum the infinite series of counterterms. The most important limitation of this procedure being the impossibility to calculate the spatial components of the boundary stress-energy tensor.

In this renormalization procedure it is fundamental that we have performed the Legendre transformation, or equivalently the addition of the boundary term in  \eqref{legendre}, canceling the divergent behavior of the gauge fields $A_1$ and $A_3$ in the on-shell action. We expect to revisit the r\^ole of the boundary-divergent fields and  to provide an extended analysis of the renormalization procedure   somewhere else.

\begin{acknowledgement}
I would like to thank Stefan Vandoren for his comments, collaboration and insight in \cite{Tarrio:2011de}. This research is supported by the Netherlands Organization for Scientic Research (NWO) under the FOM Foundation research program. I would like to thank also the Front of Galician-speaking Scientists for encouragement
\end{acknowledgement}

\end{document}